\documentclass[twocolumn,showpacs,preprintnumbers,amsmath,amssymb,,superscriptaddress]{revtex4}

\usepackage{graphicx,color}
\usepackage{dcolumn}
\usepackage{bm}
\begin{document}

\title{
Direct numerical simulations for non-Newtonian rheology of concentrated particle dispersions
}

\author{Takuya Iwashita}
 \email{iwashita@cheme.kyoto-u.ac.jp}
 \affiliation{Department of Chemical Engineering, Kyoto University, Kyoto 615-8510,Japan}
 \affiliation{CREST, Japan Science and Technology Agency , Kawaguchi 332-0012, Japan}
\author{Ryoichi Yamamoto}
 \email{ryoichi@cheme.kyoto-u.ac.jp}
\affiliation{Department of Chemical Engineering, Kyoto University, Kyoto 615-8510,Japan}
\affiliation{CREST, Japan Science and Technology Agency , Kawaguchi 332-0012, Japan}
\date{\today}

\begin{abstract}
The non-Newtonian behavior of a monodisperse concentrated dispersion of
 spherical particles was investigated using a direct numerical
 simulation method, that takes into account hydrodynamic interactions 
 and thermal fluctuations accurately. 
Simulations were performed under steady shear flow with 
 periodic boundary conditions in the three directions. 
The apparent shear viscosity of the dispersions was calculated at volume 
 fractions ranging from 0.31 to 0.56. 
Shear-thinning behavior was clearly observed at high volume fractions.  
The low- and high-limiting viscosities were then estimated from the
 apparent viscosity by fitting these data into a semi-empirical formula. 
Furthermore, the short-time motions were examined for Brownian
 particles fluctuating in concentrated dispersions, for which the fluid inertia
 plays an important role.
The mean square displacement was monitored in the
 vorticity direction at several different Peclet numbers and
 volume fractions so that the particle diffusion coefficient is
 determined from the long-time behavior of the mean square displacement. 
Finally, the relationship between the non-Newtonian viscosity of the
 dispersions and the structural relaxation of the dispersed Brownian 
 particles is examined.

\end{abstract}


\maketitle

\section{Introduction}

The links between macroscopic rheological properties and microstructures in colloidal dispersions have been extensively investigated for many systems, including dispersions of sterically-stabilized and charged-stabilized particles in host fluids \cite{ex1,ex2}.

The behavior of a monodisperse dispersion composed of solid particles immersed in a Newtonian host fluid strongly depends on the volume fraction of the dispersed particles $\Phi$ and the shear rate $\dot\gamma$. When the shear rate is zero ($\dot\gamma \rightarrow 0$), the shear viscosity of the dispersion is referred to as the zero-shear viscosity $\eta_0$. In the dilute limit ($\Phi\ll 1$), the zero shear viscosity is well approximated by Einstein's formula \cite{ex3}: 
\begin{align}
\eta_0=\eta(1 + 2.5 \Phi), \label{ein}
\end{align}
where $\eta$ is the shear viscosity of the host fluid. In a concentrated dispersion, theoretical difficulties become rather severe, since the behavior of the dispersed particles is complicated by the interactions between the particles and thermal fluctuations. In particular, the solvent-mediated many-body hydrodynamic interactions (HI) between the particles complicate the dynamical behavior. A number of experiments for concentrated dispersions have been performed to reveal the origin of the non-Newtonian behavior of these dispersions, and several semi-empirical formulas for $\eta_0$ have been proposed to characterize the experimental results. For example, the Krieger-Dougherty relationship \cite{ex4}:
\begin{align}
\eta_0=\eta\Bigl(1 - \frac{\Phi}{\Phi_m}\Bigl)^{-[\eta]\Phi_m},\label{kri}
\end{align}
where $[\eta]$ is the intrinsic viscosity and $\Phi_m$ is the packing volume fraction at which the viscosity diverges, is often used for fitting the experimental data for uniform colloidal spheres suspended in non-aqueous media. 

When a dispersion is subjected to shear ($\dot\gamma\neq 0$), the flow properties of the dispersion show a variety of non-Newtonian behaviors, such as shear-thinning and shear-thickening. These non-Newtonian behaviors are associated with the changing microstructures of the dispersion. Several physical mechanisms for these peculiar behaviors have been proposed; for example, shear-induced order-disorder transitions \cite{od0, od1,od2}, formations of dynamic clusters of the particles \cite{th1,th2}. However, a full understanding of the relationships between the rheological properties and the microstructure has not yet been obtained, despite extensive studies.

 Computer simulations are very powerful tools in the direct investigation of the dynamics of individual particles in concentrated dispersions. The Stokesian dynamics (SD) method \cite{nu1} has been widely used to measure the rheology of dispersions and provides valuable information regarding the non-Newtonian behavior of flowing dispersions \cite{SD00,SD01,SD02,SD03}. The SD method, however, is based on the Stokes approximation, which assumes that all relaxation times associated with fluid motions are short as compared with those of the particle, {\it i.e.}, $\tau_\nu \ll \tau_B$, where $\tau_\nu=\rho_f a^2/\eta$ and $\tau_B=2\rho_pa^2/9\eta$. Here, $\rho_f$ and $\rho_p$ are the density of the fluid and the particle, respectively, and $a$ is the radius of the particle. With this assumption, the HI is treated as the Ronte-Prager-Yamakawa (RPY) tensor and the lubrication correction. Furthermore, it is assumed that the relaxation time associated with the particle's inertia is zero ($\tau_B\rightarrow 0$). These approximations are valid for the motion of the particles at time scales much greater than the time scales of the relaxation of both the fluid and the particle inertia. Therefore, the short-time motion of Brownian particles over the kinematic time scale $\tau_\nu$ cannot be described by simulation methods based on the Stokesian approximation.
On the kinematic time scale, the dynamic coupling between the fluid motion and the particle motion remains strongly.

Sheared dispersions have a non-dimensional parameter that includes the particle's inertia, {\it e.g.}, the particle Reynolds number ${\rm Re_p}=\rho_p \dot\gamma a^2/\eta=9\tau_B/2\tau_{\dot\gamma}$, where $\tau_{\dot\gamma}(=1/\dot\gamma)$ represents the characteristic time due to shear. For most colloidal dispersions, the particle Reynolds number is very small. The SD method has a particle Reynolds number of zero and, therefore, cannot be applied to problems with finite particle Reynolds numbers. Typical examples include the motions of dispersions subjected to strong shear or composed of large particles; in these cases, the particle Reynolds number has a relatively large finite value.

Another problem that cannot be treated by the SD method is the short-time motion of the particle, in which the fluid inertia becomes significant even if the particle Reynolds number is very small. For example, the characteristic time scale is $\tau_\nu \sim 1 \ \mu{\text s}$ for a neutrally buoyant particle of 1 $\mu{\text m}$ in water, and ${\rm Re_p}\sim 10^{-8}$ for $\dot\gamma=0.01$. The effects of the fluid inertia appear as memory effects of the particles. For a complete understanding of these flowing dispersions, full time-dependent HIs are required.

In recent years, several numerical methods have been developed in order to accurately simulate dispersions in a variety of situations, including those described above. These methods of dispersion modeling are based on the same approach, which involves resolving the fluid motion simultaneously with the particle motion. We refer to this approach as the direct numerical simulation (DNS) approach. This approach enables us to accurately treat the full time-dependent HI. The numerical methods differ mainly in the approach used to resolve the HIs between the fluid and particle motions \cite{LB00, LB01, LB02, FP, SR01, NS01, NS02, NS03, naka1}.

In this work, we apply a direct numerical scheme based on the smoothed profile method (SPM) \cite{naka1,naka2,iwa2,iwa3,iwa1} to a monodisperse concentrated dispersion of repulsive and neutrally buoyant Brownian particles in a shear flow. In the SPM, the Navier-Stokes equation for the fluid motion is discretized on a regular grid, and the Newtonian equations for the particle motion are solved simultaneously with the fluid motion. This developed scheme accounts for thermal fluctuations, a shear flow, and memory effects. 
The SPM can reproduce the correct short-time behavior of a single Brownian particle in a shear flow on the kinematic time scale, and the numerical tests for a single Brownian particle in a shear flow were reported in \cite{iwa1}. 

Although some groups have studied flowing dispersions composed of repulsive spherical particles by using a DNS method based on the lattice Boltzmann method \cite{LB1,LB2,LB3}, all the simulations have been applied to a dispersion of non-Brownian particles and ignore the thermal fluctuations of the particles. 
Systematic analyses have not yet fully carried out
for concentrated dispersions of Brownian particles in a shear flow
at finite Reynolds numbers.

In the present study, we examine the non-Newtonian
rheology of concentrated dispersions in shear flow at finite Reynolds and
Peclet numbers using a DNS method that takes into account hydrodynamic
interactions and thermal fluctuations accurately.
Note that the simulated situations are different from those of the prevailing experiments for colloidal dispersions, since even the lowest particle Reynolds number in these simulations is several hundred times larger than those of the experiments. We first present the simulation method and the manner in which the apparent shear viscosity of the dispersions is calculated. Three-dimensional simulations are performed with periodic boundary conditions, and the non-Newtonian behavior of the shear viscosity is obtained for several volume fractions. Both the high and low shear limiting viscosities obtained from the simulations are then compared with the Krieger-Doughty relationship. 
Moreover, we investigate the short-time motions of Brownian
particles in a sheared concentrated dispersion on the kinematic time
scales. 
The mean square displacement (MSD) of Brownian particles is monitored at
several Peclet numbers and volume fractions, and the long-time diffusion
coefficient is determined from the long-time behavior of the MSD.
We also suggest a simple relationship between the non-Newtonian
viscosity of the dispersions and the structural relaxation time of the
dispersed Brownian particles.

\section{Simulation Method}

A direct numerical scheme that implements both a shear flow and thermal fluctuations is briefly explained. A more detailed explanation is given in a previous publication \cite{iwa1}. We consider a monodisperse dispersion of $N_p$ repulsive spherical particles of diameter $\sigma$ in a Newtonian host fluid. The particles interact via a truncated Lennard-Jones (LJ) potential:
\begin{align}
U_{LJ}(r) &=
\begin{cases}
 4\epsilon \Bigl[\Bigl(\frac{\sigma}{r}\Bigl)^{36} -
 \Bigl(\frac{\sigma}{r}\Bigl)^{18} + \frac{1}{4}\Bigl] & 
 (r \leq  2^{1/18}\sigma),\\
 0 & (r>2^{1/18}\sigma),\label{potential}
\end{cases}
\end{align}
where $r$ is the distance between two particles and the parameter $\epsilon$ characterizes the interaction strength. The position of the $i$th dispersed particle is ${\bm R_i}$, the translational velocity is ${\bm V_i}$, and the rotational velocity is ${\bm \Omega_i}$. The time evolution of the $\it i$th particle with mass $M_i$ and moment of inertia $\bm I_i$ is governed by Newton's equations of motion:
\begin{align}
M_i \dot {\bm V_i}&= {\bm F^H_i} + {\bm F^C_i} + {\bm G_i^V},\ \ \
\dot {\bm R_i} = {\bm V_i},\\
{\bm I_i}\cdot \dot{\bm \Omega_i} &= {\bm N^H_i} + {\bm G_i^\Omega},
\end{align}
where $\bm F^H_i$ and $\bm N^H_i$ are the hydrodynamic forces and torques exerted by the host fluid on the particle. $\bm F_i^C$ is a repulsive force arising from the potential of Eq.(\ref{potential}), which prevents the particles from overlapping. $\bm G^V_i$ and $\bm G^\Omega_i$ are random forces and torques, respectively, due to thermal fluctuations. These random fluctuations are assumed to be Markovian (white or time delta-correlated) and determine the particle temperature $T$. The procedure for determining the temperature is described in \cite{iwa2,iwa3,iwa1}. 

In the SP method, the velocity and pressure fields, $\bm v(\bm x, t)$ and $p(\bm x, t)$, are defined on three-dimensional Cartesian grids, which consist of fluid and particle domains. In order to distinguish the particle and fluid domains on the grids, a smoothed function $\phi(\bm x, t)$, which is equal to $1$ in the particle domains and $0$ in the fluid domains, is introduced. These domains are separated by a thin interfacial domain of thickness $\xi$. The system size is $[0, L_x]\times [-L_y/2, L_y/2]\times [0, L_z]$.

The time evolution of the velocity field is governed by the Navier-Stokes equation with the incompressibility condition $\nabla \cdot \bm v=0$:
\begin{align}
\rho_f(\partial_t {\bm v} + {\bm v}\cdot \nabla {\bm v}) &= \nabla \cdot\bm \sigma
 +\rho_f\phi{\bm f_p} +\rho_f \bm f^{shear},
\label{nseq}
\end{align}
where the stress tensor $\bm \sigma = -p\bm I + \eta\{\nabla \bm v + (\nabla \bm v)^T\}$, and $\bm f^{shear}(\bm x, t)$ is an external force field that is introduced to enforce a simple shear flow on the system. The flow is imposed in the $x$ direction and the external force is introduced as a constraint force so that the velocity field satisfies $v_x(y)=-\dot\gamma L_y/4$ at $y=-L_y/4$ and $v_x(y)=\dot\gamma L_y/4$ at $y=L_y/4$, where $y$ denotes the distance in the velocity gradient direction. A simple shear flow with a shear rate of $\dot\gamma$ is then approximately produced over a range from $y=-L_y/4$ to $y=L_y/4$. $\phi {\bm f_p}$ represents a body force that ensures the rigidity of the particles and the appropriate non-slip boundary conditions at the fluid/particle interface, which is further elaborated upon in reference \cite{naka1,naka2}.

{
The unit of length is taken to be the lattice spacing $\Delta$, and the
unit of time is $\tau_0=\rho_f\Delta^2/\eta$. Unless otherwise stated, we set
$\Delta=1$, $\tau_0=1$, $\eta=1$, $\rho_f=1$, $\rho_p=1$, $a=4$, $\epsilon=1$, and
$\sigma=8$.
Assuming dispersions of neutrally buoyant particles of
radius $1\mu$m in water at room temperature, 
our unit length $\Delta$ and time $\tau_0$ correspond to 
$0.25\mu$m and $0.0625\mu$s, respectively
The second-order Runge-Kutta
algorithm is used to integrate the Newtonian equations. The
Navier-Stokes equation is discretized with a Fourier spectral scheme in
space and with a second-order Runge-Kutta scheme in time. The
discretized time step is $h=0.07145$. Although this value is choosen
from the stability condition of the Navier-Stokes equation, 
it can be used safely for the particle's equations motion because
$h$ is much smaller than the Lennard-Jones time unit
$\tau_M=(M_i\sigma^2/\epsilon)^{1/2}\simeq131$.
}
The phase diagram of the present (36:18) LJ system depends weakly on the system temperature $T$ as well as the volume fraction $\Phi$ of the particles. To avoid the delicate issue of crystallization, all the simulations in the present paper were carried out at $T$ and $\Phi$ with which the system is in amorphous states.

To measure the rheological properties of the particle dispersion in
shear flow, we can calculate the apparent stress $\bm \sigma^{app}$ of
the dispersions in the following manner. The momentum equation for the dispersion is formally written as: 
\begin{align}
\frac{d}{dt}(\rho_t\bm v) &= \nabla \cdot \bm \sigma^{dis} + \rho_t\bm f^{shear},\\
\rho_t &= (1-\phi)\rho_f + \phi\rho_p,
\end{align}
where $\bm \sigma^{dis}$ denotes the stress tensor of the dispersion
including the inertia, the pressure and the viscous terms. 
The full stress tensor $\bm s$ of the flowing dispersion is then defined by
introducing a convective momentum-flux tensor explicitly as:
\begin{align}
{\bf s}= {\bm \sigma^{dis}} - \rho_t {\bm v}{\bm v},
\end{align}
where $\rho_t {\bm v}{\bm v}$ represents momentum transport by the bulk flow of the dispersion.

Although  $\bm s$ cannot be calculated directly, we can obtain the apparent stress $\bm \sigma^{app}$ by using the local stress $\bm s$:  
\begin{align}
\bm \sigma^{app} &= \frac{1}{V}\int d\bm x \bm s\\
                 &= \frac{1}{V}\int d\bm x \Bigl[[\nabla \cdot(\bm s\bm x)]^T - \bm x\nabla \cdot \bm s\Bigl]\label{FM} \\ 
                 &= \frac{1}{V}\int d\bm x [-\bm x \nabla \cdot \bm s]\\                 &= \frac{1}{V}\int d\bm x \Bigl[\bm x \Bigl(\rho_t \bm f^{shear}- \frac{\partial}{\partial t}(\rho_t \bm v)\Bigl)\Bigl]\\
                 &= \frac{1}{V}\int d\bm x \bm x \rho_t \bm f^{shear} - \frac{1}{V}\int d\bm x \bm x\frac{\partial}{\partial t}(\rho_t \bm v),
\end{align}
with a volume $V=L_xL_yL_z$. In the derivation of Eq.(\ref{FM}), we use an second rank identity, $\bm s =[\nabla \cdot(\bm s \bm x)]^T - \bm x\nabla \cdot \bm s$. In the steady state, 
\begin{align}
\Bigl\langle \frac{\partial }{\partial t}(\rho_t \bm v) \Bigl\rangle_{t}=0,
\end{align}
where $\langle \rangle_t$ denotes time averaging over the steady state. The time-averaged apparent shear stress of the dispersion can then be written as: 
\begin{align}
\langle \bm \sigma^{app} \rangle_{t} = \frac{1}{V}\Bigl\langle\int d\bm x \bm x  \rho_t \bm f^{shear}\Bigl\rangle_{t}.
\end{align} 
{Then one can obtain the apparent shear stress under steady
shear flow from the external force $\bm f^{shear}$ imposed in the
Navire-Stokes equation.}

\section{Results and discussion}

{Simulations were performed in a three-dimensional cubic box, 
whose side length is $L$, with periodic boundary conditions. 
Most present simulations were done with $L=64$, where the number of particles
are $N_p=300$, $400$, $450$, $500$, and $550$ for
$\Phi=0.31$, $0.41$, $0.46$, $0.51$, and $0.56$, respectively.}
The index of axis $x$, $y$, and $z$ represent the flow, velocity gradient, and vorticity directions, respectively.
For a spherical particle, $a=4$, $\xi=2$. The temperature was determined by equilibrium calculations before the shear flow was imposed. The temperature is $k_BT=7$. The initial configuration of the particles is set to be a random distribution. A large number of simulations were performed for volume fractions of $0.31\leq \Phi\leq 0.56$ and shear rates of $5 \times 10^{-5}\leq \dot\gamma\leq 0.1$. These systems have particle Reynolds numbers $\rm Re_p$ ranging from $8\times10^{-4}$ to 1.6.

For dispersions composed of Brownian particles in the steady state, the apparent shear viscosity of the dispersion is defined as: 
\begin{align}
\eta^{app}= \frac{\langle \sigma^{app}_{xy}\rangle_{t}}{\dot\gamma},
\end{align}
where $\sigma^{app}_{xy}$ denotes the $xy$-components of $\bm \sigma^{app}$. 
{To examine the system size effects in the present simulations, we
calculated the apparent shear viscosity for three different 
system sizes for a constant $\Phi=0.41$, {\it i.e.}, 
$L=32$ with $N_p=50$, $L=128$ with $N_p=3200$, 
and $L=256$ with $N_p=25600$.
We found only negligibly small differences among the values of 
the apparent shear viscosity calculated for the three systems
within a statistical error.
}

\begin{figure}[h]
\begin{center}
\includegraphics[scale=0.9]
{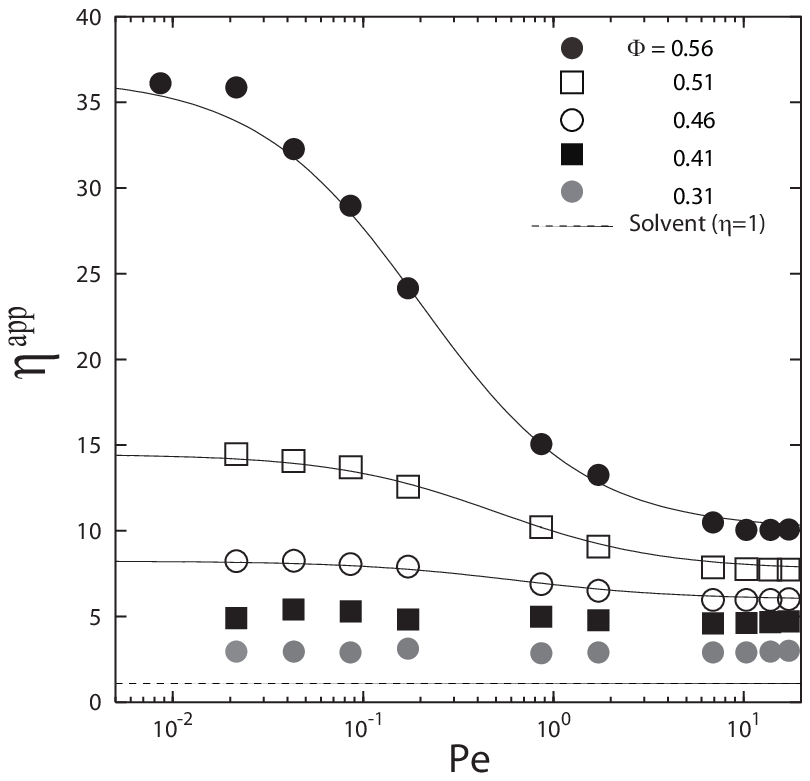}
\end{center}
\caption{\label{visco_peclet}{
The Peclet number ${\rm Pe}= 6\pi\eta a^3 \dot\gamma /k_BT$
dependence of the apparent shear viscosity of the dispersion for several volume fractions. 
The solid lines represents $\eta_f=\eta_{\infty} +
 (\eta_0-\eta_\infty)/(1+b^{-1}(\Phi){\rm Pe})$ for $\Phi\geq 0.46$,
 where $\eta_0$ is the low shear limiting viscosity, $\eta_{\infty}$ is
 the high shear limiting viscosity, and ${\it b}(\Phi)$ is a fitting
 parameter, with $b=0.20$, $0.50$, and $0.63$ for $\Phi=0.56$, $0.51$,
 and $0.46$, respectively. The dashed line represents the shear viscosity of the host fluid.}
}
\end{figure}

Figure \ref{visco_peclet} shows the dependence of the apparent shear viscosity on the Peclet number for several volume fractions. Here, the Peclet number is defined as ${\rm Pe}=6\pi\eta a^3 \dot\gamma/k_BT$. 
For the lowest concentration ($\Phi=0.31$), the viscosities remain nearly constant and exhibit Newtonian behavior. For higher concentrations ($\Phi\geq 0.41$), the dispersions show non-Newtonian behavior. As $\rm Pe$ increases, the shear-thinning behavior is clearly observed from the higher plateau region for $\rm Pe$ of order $10^{-2}$ to a lower plateau region for $\rm Pe$ of about 10. From both of the plateau values of the viscosity curve, we can obtain the low shear limiting viscosity (identified as $\eta_0$) and the high shear limiting viscosity $\eta_{\infty}$ for each volume fraction.

In order to evaluate $\eta_0$ and $\eta_{\infty}$, we fit our simulation
data into the following simple empirical function of $\rm Pe$ and $\Phi$:
\begin{align}
\eta_{f}({\rm Pe},\Phi) = \eta_{\infty} + \bigg(\frac{\eta_0-\eta_{\infty} }{1+b^{-1}(\Phi){\rm Pe}}\bigg),
\end{align}
where $\rm {\it b}(\Phi)$ is a fitting parameter.
This empirical function is plotted in Fig. \ref{visco_peclet} with solid
lines for $\Phi\geq 0.46$.
One finds that it coincides with the simulation data reasonably well.
It is also seen that the onset of shear thinning appears at smaller 
Peclet numbers with increasing volume fraction.
We note that the particles are randomly distributed in the
dispersions in the present simulations, {\it i.e.} all the viscosity
data are taken in situations at which no shear-induced crystallization
occurs.

\begin{figure}[h]
\begin{center}
\includegraphics[scale=0.9]
{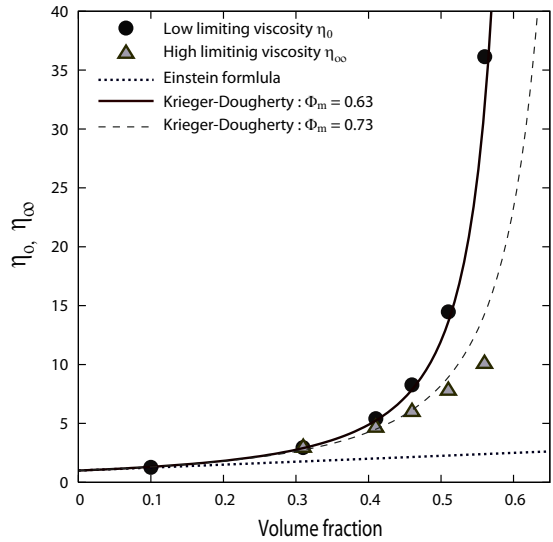}
\end{center}
\caption{\label{visco_volume}{
Volume fraction dependence of the low shear limiting viscosity $\eta_0$ and the high shear limiting viscosity $\eta_{\infty}$ of the dispersions. The solid line represents the Krieger-Dougherty relationship with $\Phi_m=0.63$ for $\eta_0$ and the dashed line $\Phi_m=0.73$ for $\eta_{\infty}$. The dotted line represents Einsteinfs formula of Eq.(\ref{ein}).
}}
\end{figure}

Figure \ref{visco_volume} shows the dependence of the low shear limiting viscosity $\eta_0$ and the high shear limiting viscosity $\eta_{\infty}$ on the volume fraction. Both shear limiting viscosities increase monotonically with the volume fraction. The simulation results for both of the shear limiting viscosities agree well with  the semi-empirical relations of Eq.(\ref{kri}) by Krieger-Dougherty, $\eta_{0,\infty}=\eta(1- \Phi/\Phi_m)^{-2.5\Phi_m}$ where $\Phi_m=0.63$ for $\eta_0$ and $\Phi_m=0.73$ for $\eta_{\infty}$. The dotted line represents Einstein' s formula of Eq.(\ref{ein}). 

The fluid and particle inertia contribution to the apparent shear stress can be written as 
\begin{align}
\sigma^{inertia}_{xy} = \frac{1}{V} \int d\bm x \rho_t v_x v_y,
\end{align}
which is similar to the Reynolds stress of a uniform fluid in turbulence. This stress represents the strength of the hydrodynamic instability. 
The inertia contribution to the viscosity, $\eta^{inertia}=\langle \sigma_{xy}^{inertia}\rangle_t/\dot\gamma$, 
is about two order of magnitude smaller than the apparent shear viscosity.
We conclude that the inertia contribution is negligible 
in the range $8 \times 10^{-4} \leq {\rm Re_p}\leq 1.6$.

We next examine the dynamical motion of Brownian particles in a shear flow. We analyze the mean square displacement (MSD) in the vorticity direction ($z$) for the Brownian particles,
\begin{align}
\langle [\Delta R^z(t)]^2\rangle = \frac{1}{N_p}\sum_{j=1}^{N_p}\langle |R_j^z(t) -R_j^z(0)|^2 \rangle.
\end{align} 

\begin{figure}[h]
\begin{center}
\includegraphics[scale=0.8]{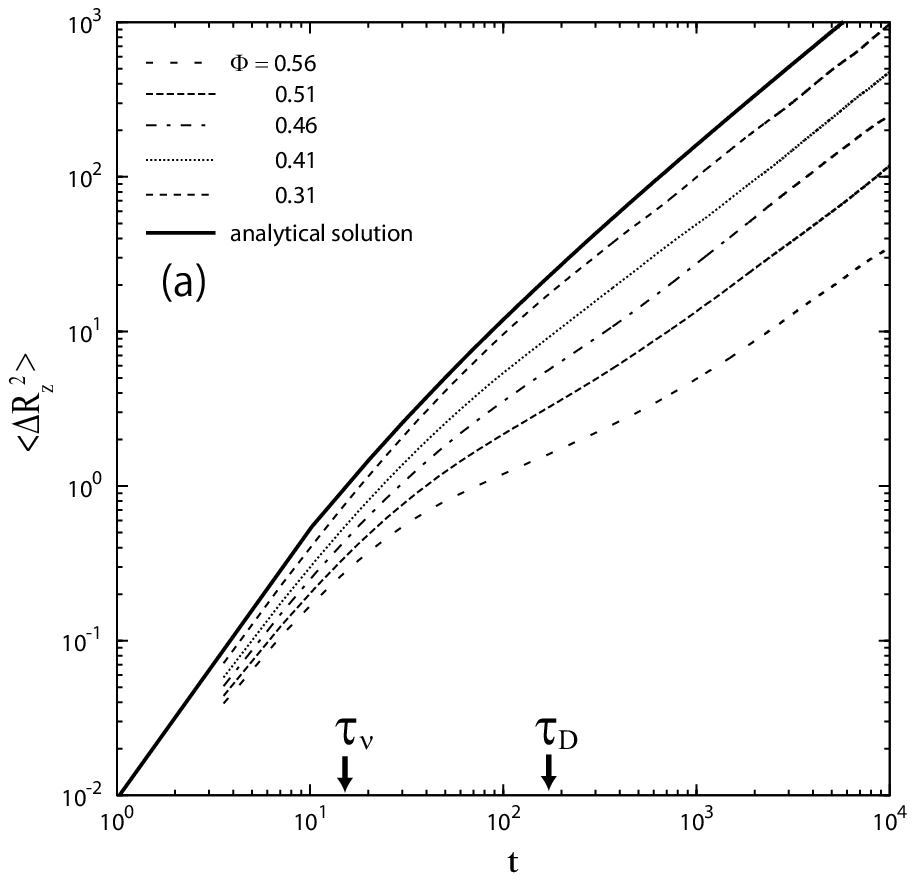}
\includegraphics[scale=0.8]{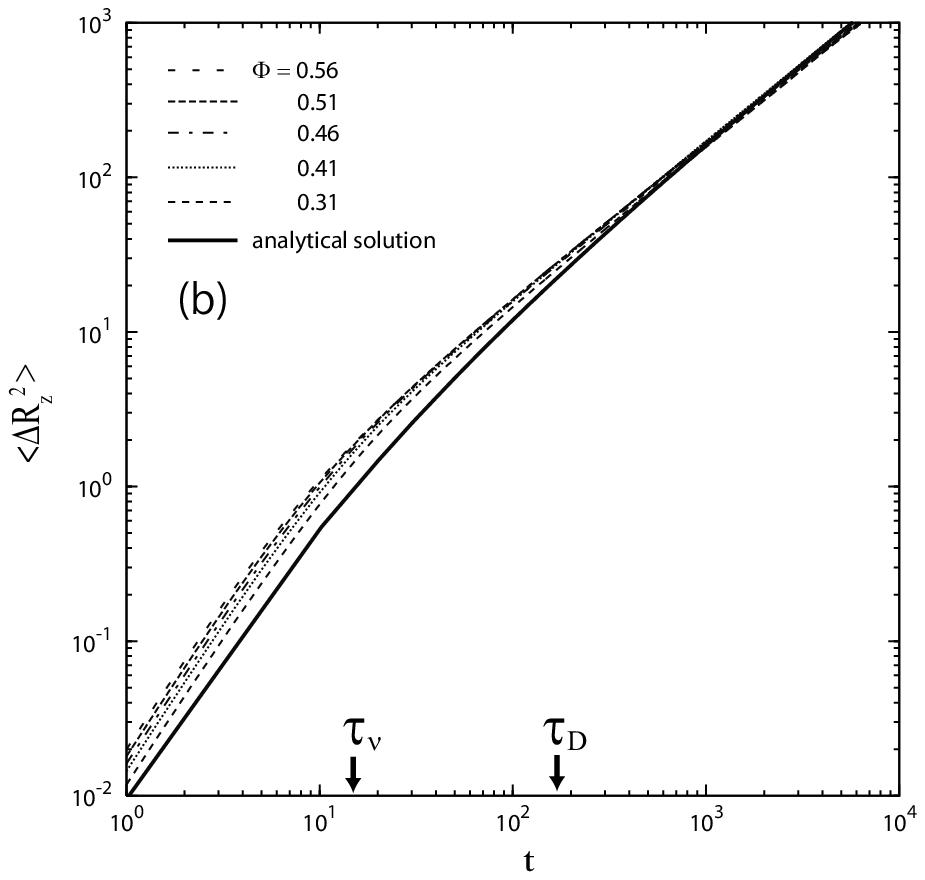}
\end{center}
\caption{\label{msd0}{
The mean square displacement in the vorticity direction for several 
volume fractions (a) at the lowest Peclet number 
(${\rm Pe}=0.0086$ at $\Phi=0.56$ and ${\rm Pe}=0.022$ for $\Phi\leq
 0.46$) and (b) at the highest Peclet number (${\rm Pe}=17.2$).
The solid line represents the analytical solution for 
the generalized Langevin equation of a free Brownian particle \cite{GLE}. 
The arrows indicate the kinematic time $\tau_\nu=\rho_fa^2/\eta=16$ 
and the diffusion time $\tau_D=a^2/D_0\sim 172$.
}
}
\end{figure}

Figure \ref{msd0} (a) shows the MSD of the dispersion with the lowest Peclet number for each volume fraction. These results are considered to be the MSD at thermal equilibrium. As the volume fraction is increased, the dynamical behavior of the MSD varies greatly, as compared with that of the hydrodynamic analytical solution for a single Brownian particle in a shear flow \cite{GLE}. 
The analytical solution is derived from the generalized
Langevin equation with memory effects.
With increasing volume fraction, MSD increases more slowly with time.
For $\Phi=0.56$, we found that a plateau region starts to appear
around a time scale of order $10^2$, which is slightly greater 
than the kinematic time $\tau_\nu$. 
This is a typical behavior of colloidal dispersions in glassy states, {\it e.g.}, colloid glasses.

Figure \ref{msd0} (b) shows the MSD of the dispersion at the highest
Peclet number for each volume fraction. Since $\rm Pe \simeq 17.2$ here,
the shear force is much stronger than the thermal force.
The volume fraction dependence of the MSD in this figure is opposite to 
the previous case at the lowest Peclet number shown in Fig. \ref{msd0} (a).
At short times, the MSD grows more rapidly in time with increasing
volume fraction because HIs between particles are more enhanced at
higher volume fraction. 
At later times, 
the MSDs tend to exhibit diffusive motions with a volume fraction
independent diffusion coefficient.

{We then calculated the long-time diffusion coefficient $D^{sim}_z$ in the vorticity
direction via
\begin{align}
D^{sim}_z= \lim_{t\rightarrow \infty} \frac{1}{2t} \langle [\Delta R^z(t)]^2\rangle
\end{align}
and examined the system size effects in $D^{sim}_z$.
In contrast to the negligibly small system size effects observed in
the apparent shear viscosity $\eta^{app}$, 
$D^{sim}_z$ shows notable increase with increasing system size for all volume
fractions.
It is, however, confirmed that the diffusion coefficients behave as 
\begin{equation}
D_z^{sim}(\Phi, L^{-1}) = D_z(\Phi) - \xi k_BT/6\pi\eta L
\end{equation} 
with $\xi\sim 2.83$, where $D_z(\Phi)$ represents the value of diffusion coefficient
extrapolated for $L\rightarrow\infty$.
The system size effects observed in the present simulations are
essentially the same as those reported in earlier papers 
\cite{dunweg,system01,system02}. 
We calculated $D_z(\Phi)$ from extrapolation of our simulation data $D_z(\Phi, L)$ with
finite $L$.
Figure 4 shows the volume fraction and the Peclet number dependencies 
of $D_z(\Phi)$ normalized by $D_0=k_BT/6\pi\eta a$.
}
For the low Peclet numbers, the diffusion coefficient decreases with increasing volume fraction, and this dependence is similar to that of the dispersion at thermal equilibrium. 
On the other hand, for the highest Peclet number, the volume fraction dependence of $D_z$ is almost constant. We can see that the diffusion coefficients at high volume fractions are strongly affected by shear. The dynamics of the Brownian particles differs considerably, depending on whether the thermal or shear force is dominant. The volume fraction dependence is in qualitative agreement with the numerical results achieved based on the SD method, which is valid for long time scales \cite{SD00}.

Finally, the relationship between the non-Newtonian viscosity of the
 dispersions and the structural relaxation of the dispersed Brownian 
 particles is examined.
The structural relaxation time is defined as 
\begin{align}
\tau_p(\Phi)=a^2/D_0(\Phi)
\end{align}
where $D_0(\Phi)$ is the diffusion coefficient for each volume fraction
at zero shear. This relaxation time represents a time needed for a particle
to diffuse away a distance comparable to its radius. 
Figure \ref{diffusion1} shows 
$(\eta^{app} - \eta_\infty)/(\eta_0 -\eta_\infty)$ 
versus $\tau_p\dot\gamma$ for $\Phi\geq 0.46$. 
Here, we used $D_z (\Phi)$ at the lowest Peclet number as $D_0 (\Phi)$, 
and $\tau_p\dot\gamma$ can be understood as a reduced 
Peclet number ${\rm Pe} (\Phi)=\dot\gamma a^2/D_0 (\Phi)$. 
One can see that the data lies on a single master curve
$1/(1+Ax)$ fairly well, where $x=\tau_p\dot\gamma$ and $A$ is a fitting
parameter. 
The shear-thinning starts at around $\tau_p\dot\gamma\simeq 1$. 
This behavior is very similar to the non-Newton rheology of supercooled
liquids \cite{glassy}.

We confirmed that the present numerical method of introducing the thermal fluctuation successfully reproduces the fluctuation dissipation theorem for time scales longer than the so-called the Brownian time \cite{iwa2,iwa3}. Although it was confirmed also that the present method works quite well for the volume fractions $0 \le \Phi \le 0.56$ considered in the present study, further careful tests must be needed for highly concentrated dispersions, where the fluctuation of a tagged particle tends to correlate with motions of surrounding particles.

\section{Conclusions}
We investigated the rheological properties of monodisperse concentrated dispersions with repulsive spherical particles by using a DNS method that accounted for a shear flow and thermal fluctuations. 
Three-dimensional simulations were performed at Peclet numbers ranging 
from $0.043$ to $17.2$ and 
at particle Reynolds numbers from $8\times 10^{-4}$ to $1.6$.
The apparent viscosity for $\Phi=0.31$ is almost constant over the
Peclet number change. 
For $\Phi\geq 0.41$, the viscosities decrease from the plateau region at the low Peclet number to the plateau region at the high Peclet number.
From the viscosity versus Peclet number curves, we can obtain both the low and high limiting viscosities, and these results are in good agreement with the Krieger-Doughty relationship.
The inertia contribution to the apparent shear viscosity is very small
throughout the entire range of Peclet numbers and volume fractions
examined in the present study.
As the volume fraction increases, the behavior of the MSD in the vorticity direction deviates from the analytical solution for a single Brownian particle in a shear flow.
For the lowest Peclet number, the MSD develops more slowly in time with
increasing volume fraction. At $\Phi=0.56$, an onset of glassy dynamics 
was observed. 
On the other hand, for the highest Peclet number, the MSD develops more
rapidly at short times  with increasing volume fraction.
Finally, the volume fraction dependence disappears at long times.
The diffusion coefficient was calculated from the long-time behavior 
of the MSD in the vorticity direction for different Peclet numbers 
and volume fractions.
For the lowest Peclet number, the diffusion coefficient decreases with
increasing volume fraction, while it is almost constant over the volume
fraction change for the highest Peclet number. 
We estimated the structural relaxation time $\tau_p$ from the diffusion
coefficient at the lowest Peclet number.
The present non-Newtonian viscosity data agrees well with a simple
scaling function 
$(\eta^{app} - \eta_0)/(\eta_\infty- \eta_0) = 1/(1+A\tau_p\dot\gamma)$
for $\Phi\geq 0.46$, similar to the non-Newtonian viscosity of 
a model supercooled liquid \cite{glassy}.

\newpage
\widetext

\begin{figure}[h]
\begin{center}
\includegraphics[scale=0.8]{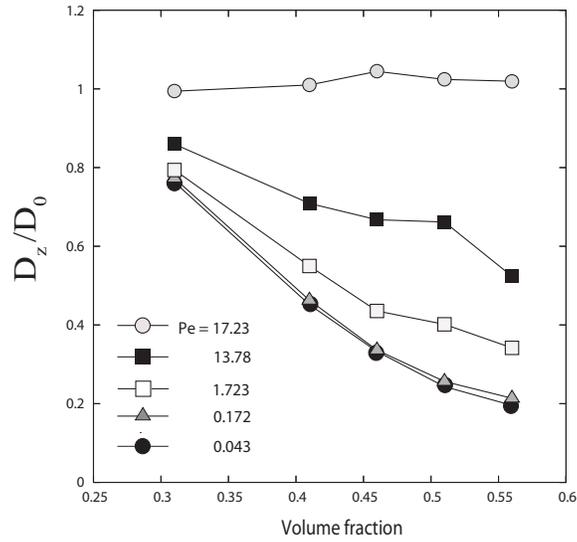}
\end{center}
\caption{\label{diffusion}{
The long-time diffusion coefficient in the vorticity direction for different Peclet numbers and volume fractions. The diffusion coefficient is scaled by $D_0=k_BT/6\pi\eta a$.
}}
\end{figure}

\begin{figure}[h]
\begin{center}
\includegraphics[scale=0.8]{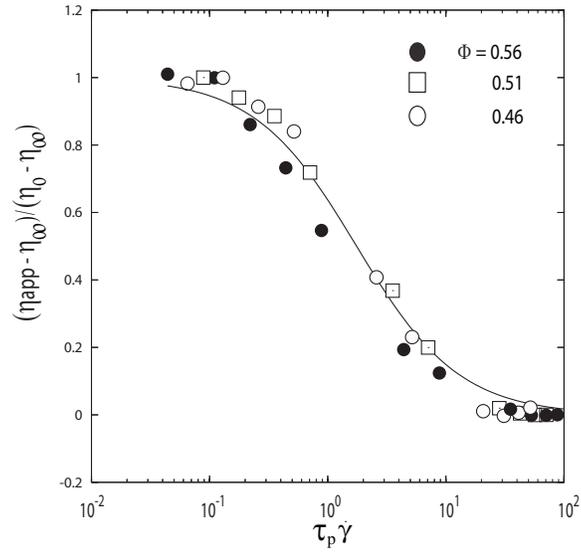}
\end{center}
\caption{\label{diffusion1}{
Scaled plot of $(\eta^{app} - \eta_{\infty})/(\eta_0 - \eta_{\infty})$ against reduced Peclet number $\tau_p \dot\gamma$ for $\Phi \geq 0.46$. Line is $1/(1+ A\tau_p\dot\gamma)$ where A=0.57.
}}
\end{figure}

\end{document}